\documentclass[aps,prb,preprint,groupedaddress]{revtex4-1}
\usepackage{graphicx}

\begin{document}


\title{Achieving balanced open circuit voltage and short circuit current by tuning the interfacial energetics in organic bulk heterojunction solar
cells: A drift-diffusion simulation}


\author{Wenchao Yang}
\email[]{wchy@fudan.edu.cn}
\affiliation{Key Laboratory of Microelectronics and Energy of Henan, 
School of Physics and Electronic Engineering, Xinyang Normal University, Xinyang, 464000, China}


\date{\today}

\begin{abstract}
In organic bulk heterojunction solar cells, the donor/acceptor interfacial energy offset ($\Delta E$) is found to provide the driving force for efficient charge separation which gives rise to high short circuit current density ($J_\mathrm{sc}$), but a high $\Delta E$ inevitably undermines the open circuit voltage ($V_\mathrm{oc}$). In this paper, employing the device model method we calculated the steady state current density-voltage ($J-V$) and the $J_\mathrm{sc}-\Delta E$ curves under two different charge separation mechanisms to investigate the optimum driving force required for achieving sizable $V_\mathrm{oc}$ and $J_\mathrm{sc}$ simultaneously. Under the Marcus charge transfer mechanism, with the increased $\Delta E$ the Jsc increases rapidly for $\Delta E\leq 0.2$ eV, and then maintains a nearly constant value before decreasing at the Marcus inverted region, which is due to the accumulation of undissociated excitons within their lifetime and is beneficial for obtaining a sizable $J_\mathrm{sc}$ under a $\Delta E$ much smaller than the reorganization energy $\lambda$. With inclusion of both the electron and the hole transfer pathways of different respective $\lambda$'s into the device model, the experimentally measured $J-V$ curves for donor/acceptor blend with different $\Delta E$'s can be reproduced. For the coherent charge transfer mechanism in which the driving force act as the energy window of accessible charge separated states, with two typical types of density of states for the charge transfer excitons, it is shown that the highest $J_\mathrm{sc}$ can also be achieved under a small $\Delta E$ of 0.2\,eV if the high-lying delocalized states are harvested in high proportion. This work demonstrates the existence of the optimum driving force of 0.2\,eV and provides some guidelines for engineering the interfacial energetics to achieve the high balanced $J_\mathrm{sc}$ and $V_\mathrm{oc}$.
\end{abstract}


\maketitle

\section{Introduction}
In organic bulk heterojunction solar cells, the randomly oriented donor/acceptor (D/A) interfaces are generally employed for converting the photogenerated excitons into free charge carriers\cite{deibel,hains,clark,gunes,hedley}. There have been plenty of experimental and theoretical works devoted to the investigation of the various interfacial properties, which demonstrated that the performance and stability of the devices are largely determined by the interfacial donor/acceptor morphology\cite{bavel,huang,szarko,nuzzo}, the interfacial molecular orientation and aggregation\cite{barry,miller,fu,chen,ryno,ran,ndjawa}, since they have significant impacts on the interfacial energetics\cite{poelking,guo,fazzi}. In particular, the interfacial energetics plays the central role on the exciton dissociation and charge generation\cite{clark,hedley}, and thus attracted most of the research attention among all the interfacial properties. It is expected that through investigating the working principles of the interfacial energetics people can optimize it and fabricate high efficiency photovoltaic devices. However, the interfacial energetics is rather complicated, which not only involves the molecular frontier orbitals of the donor and acceptor materials, but also the tightly-bounded singlet excitonic states, the loosely-bounded charge transfer (CT) states and the charge separated (CS) states upon photo-excitation\cite{ohkita,tvingstedt,deibel2,arndt,grancini,jailaubekov,nuzzo}. Moreover, each type of these excited states consists of many energy levels forming a manifold\cite{clark,barker,gerhard,ti}. Up to now, the intricate interactions among the states and their effects on the charge separation processes remain hotly debated.

Macroscopically, the interfacial energetics is revealed to have direct impacts on both the short circuit current density ($J_{sc}$) and the open circuit voltage ($V_{oc}$) of the devices. For the $V_{oc}$, its upper limit is basically determined by the CT states energy and their disordered effect\cite{veldman,vandewal,burke,collins,zou,guan}; while for the $J_{sc}$, it is experimentally demonstrated that a finite lowest unoccupied molecular orbital (LUMO) or highest occupied molecular orbital (HOMO) level offset across the interface is indispensable for obtaining sizable photocurrent\cite{ohkita,dnuzzo,hoke,hendriks}. The measured current density-voltage ($J-V$) characteristics for devices with a fixed donor material and different fullerene acceptor materials suggest that, as the acceptor with higher LUMO level is employed, the $V_{oc}$ becomes larger due to the increased effective band gap, whereas the $J_{sc}$ decreases significantly. Especially for the recently popular fullerene-based acceptor of ICBA, when it is blended with P3HT as the photoactive layer, the interfacial LUMO offset is smaller than 0.05eV, and the corresponding $J-V$ curve exhibits a $V_{oc}$ of over 1\,V but an extremely small $J_{sc}$, representing poor charge generation efficiency\cite{dnuzzo,hoke}. For the hole transfer processes, the required HOMO energy offset is found to be even 0.3\,eV higher than the driving force for electron transfer\cite{hendriks}. Thus, the interfacial LUMO (HOMO) offset is believed to play important roles on charge separation and is usually referred as the driving force for charge transfer and separation in literature\cite{rand,clark,ohkita,shoaee, ward,coffey,dimitrov,wright,jakowetz}. More rigorously, the driving force can be defined as the difference between the effective band gap and the CT state energy\cite{jakowetz}.

According to the different scenarios proposed to explain how the exciton dissociation and charge separation process proceed at the donor/acceptor interface, the possible roles of the driving force could be the following three folds. First of all, the charge transfer at the interface may be a non-adiabatic process which involves a relatively large reorganization energy. The energy level offset provides the free energy for carriers to reach the intersection of potential energy surfaces and achieve resonant charge transfer, as described by the traditional Marcus theory\cite{ljakoster,zhang,wright,volpi}. This mechanism has been demonstrated by the measurement of photo-carrier yield for a series of acceptors\cite{coffey}. Secondly, the driving force may provide the kinetic energy required for the electron-hole polaron pairs to escape from their mutual Coulomb attractive potential, which is the so-called hot CT state dissociation\cite{clark,ohkita,shoaee,murthy,grancini,jailaubekov,schulze,fuzzi}. Thirdly, employing the pump-push-photocurrent measurements on the free carrier generation efficiency, Bakulin et al found that there exists a band of delocalized high-lying CT states or some vibrational modes which can facilitate the coherent transport of charge carriers on these states to achieve full separation, while those charge carriers on the low-lying CT states generally recombine geminately and do not contribute to the photocurrent\cite{bakulin1,bakulin2,jakowetz}. Due to this coherent charge separation mechanism the finite driving force seems unnecessary\cite{chner,kaake,whaley}. However, with the increased energy level offset, more delocalized states become accessible for the ballistic or coherent transfer of charge carriers, so that the thus measured charge generation rate still exhibits a weak dependence on the LUMO energy level offset\cite{jakowetz}. On the other hand, the impacts of donor: acceptor ratio is much stronger in this case, because the high proportion of fullerene acceptor material will spontaneously aggregates and forms crystalline phases, which give rise to much more delocalized electronic states\cite{jakowetz,savoire,tamura,nan}.

Each of the charge generation mechanisms can partly explain the experimental phenomena and it is highly controversial that which one is the dominant. Since most of the measurements are done under the transient pulsed luminescence conditions, it remains unclear to what extent the steady state performance of devices is limited by the charge generation rates under the incoherent (Marcus) or coherent mechanisms. Moreover, as mentioned above, there is always a tradeoff between the $J_{sc}$ and $V_{oc}$ under a specific value of the driving force, and increasing the driving force to boost the free carrier generation will inevitably lead to the decreased $V_{oc}$\cite{rand,ohkita}. Thus people need to find the minimum driving force required for efficient charge separation to avoid sacrificing the $V_{oc}$ too much. Actually this has been realized in some non-fullerene acceptor solar cells, where the high and balanced $J_{sc}$ and $V_{oc}$ can be reached simultaneously\cite{liu,baran}. But the theoretical explanation for this desirable effect is still lacking.

In this paper, we employ the macroscopic device model simulation to investigate the effect of the driving force on the final device performance, especially the $J_{sc}$ which was less intensively studied than the properties of $V_{oc}$ in literature. The interfacial energy offset are incorporated into the device model both through its impacts on the effective band gap and on the exciton dissociation rate or proportion to calculate the $J-V$ curves under different driving forces. The investigations are done on the theoretical frameworks of the incoherent and the coherent charge separation mechanisms. It is found that both of them can give rise to $J-V$ curves similar to the experimentally measured ones. Moreover, there indeed exists an optimum driving force of 0.2\,eV or so for obtaining balanced $J_{sc}$'s and $V_{oc}$'s. Under the steady state, with the incoherent dissociation mechanism the relatively large $J_{sc}$ can be achieved in a broad range of the interfacial LUMO offset so that it could be restricted to a much less value than the reorganization energy; while with the coherent mechanism, the denser is the distribution of the delocalized CT states above the acceptor LUMO level, the higher is the $J_{sc}$ under small driving forces. The results are consistent with the finding that the efficient charge separation can be achieved under small driving forces\cite{lee,heeger}, and may also provide clues for the design and preparation of the organic donor/acceptor with optimized interfacial energetics to fabricate devices of high power conversion efficiency. In Sec.\ref{method}, we describe the model we used in simulation, and in Sec.\ref{results}, the simulated J-V curves for the incoherent and the coherent dissociation mechanisms are shown, respectively, and the variation of $J_{sc}$ under different driving forces is discussed in detail. Finally, we give the conclusions in Sec.\ref{conclusion}.

\section{Theoretical device modeling method}\label{method}
The one dimensional device model provide a straightforward method to calculate the device operating parameters under the influences of various microscopic electronic processes\cite{smith,blom}. For the bulk heterojunction devices, the active layer in which the donor and the acceptor phases interpenetrate with each other and form percolating pathways for charge transport is considered as a homogeneous medium. Although the interfacial morphology cannot be taken into account in the model, for finely-mixed donor/acceptor phases this assumption is valid from a macroscopic point of view. In order to produce free charge carriers, the photo-generated singlet excitons must experience two successive dissociation steps. In the first one, the exciton diffuses to the donor/acceptor interfaces and transfer their electrons from the donor phase to the acceptor phase while leaving the holes in the donor phase, forming CT states on the interfaces\cite{clark}. The exciton dynamics is described by the following continuity equation
 \begin{equation}\label{exciton}
    \frac{\partial X}{\partial t}=D_X\frac{\partial^2 X}{\partial x^2}-\frac{X}{\tau}-k_\mathrm{PET} X + G,
 \end{equation}
where $X$ is the exciton concentration. The terms on the right-hand side of Eq.~(\ref{exciton}) represents the diffusion, the radiative and non-radiative decay, the dissociation and the photo-generation processes, respectively, with $D_X$ the diffusion coefficient, $\tau$ the lifetime, $k_\mathrm{PET}$ the photo-electron transfer rate and $G$ the optical generation rate. This ultrafast electron transfer process is nonadiabatic and its rate is given by the Marcus theory:\cite{clark}
 \begin{equation}\label{ket}
    k_\mathrm{PET}=\frac{2\pi}{\hbar\sqrt{4\pi \lambda kT}}V^2\exp\left(-\frac{(\triangle G+\lambda)^2}{4\lambda kT}\right),
 \end{equation}
in which the $V$ stands for the electronic coupling between the donor and acceptor molecules; the $\lambda$ represents the reorganization energy; and the $\triangle G$ is the free energy. In the context of charge transfer at the donor/acceptor heterojunction, the $\triangle G$ is actually equal to the interfacial energy offset. The value of $k_\mathrm{PET}$ is mainly dominated by the exponential factor on the right side of Eq.~(\ref{ket}), and the corresponding prefactor is assumed to be a constant of $k_0$, which may also represents the coherent (ballistic) charge transfer rate. The coherent charge transfer mechanism arises from the delocalized CT states and its modeling method is postponed to the Sec.\ref{results} for compactness.

In the incoherent charge separation mechanism, only certain proportion of the thus produced CT states can dissociate and generate free charge carriers, while the others recombine geminately to the ground state. According to the Onsager-Braun theory, the proportion of the successfully dissociated CT states $P(E)$ is mainly dependent on temperature, the electric field strength, and the CT states binding energy. With the approximate form of\cite{clark}
\begin{equation}\label{pe}
    P(E)=\exp\left(-\frac{e^2}{4\pi\varepsilon_0\varepsilon kT a}\right)\left(1+\frac{e^3}{8\pi\varepsilon_0\varepsilon (kT)^2}E\right),
\end{equation}
it is incorporated into the free carrier generation rate. Now the continuity equations for electrons and holes can be written as:
 \begin{eqnarray}
   \frac{\partial p}{\partial t} &=& -\frac{1}{e}\frac{\partial J_p}{\partial x}+P(E)k_\mathrm{PET} X-R, \label{pt} \\
   \frac{\partial n}{\partial t} &=& \frac{1}{e}\frac{\partial J_n}{\partial x}+P(E)k_\mathrm{PET} X-R.  \label{nt}
 \end{eqnarray}
The electron (hole) current $J_n$ ($J_p$) has the common drift-diffusion form\cite{blom}, with the Einstein's relation being assumed. At the two ends of the device, the $J_n$ and $J_p$ are defined as the respective net surface recombination currents\cite{sandberg}, which consist of the boundary conditions for Eqs~(\ref{pt},\ref{nt}). The bimolecular recombination rate
$$R=\zeta \frac{e(\mu_n+\mu_p)}{\varepsilon_0\varepsilon}(np-n_i^2)$$ where $\zeta$ is the reduction factor with respect to the Langevin bimolecular recombination rate\cite{burke}.

The internal electric field $E(x)$ obeys the ordinary Poisson's equation
\begin{equation}\label{poisson}
    \frac{\partial E}{\partial x}=\frac{e}{\varepsilon_0\varepsilon}(p-n)
\end{equation}
with the constraint that
\begin{equation}\label{cons}
    \int_0^L E(x)dx=V_\mathrm{ext}-(E_g-\phi_p-\phi_n)/e ,
\end{equation}
in which $V_\mathrm{ext}$ is the externally applied bias voltage, $E_g$ is the effective band gap, and $\phi_p(\phi_n)$ is the hole (electron) injection barrier at the anode (cathode), namely the effective voltage drop across the device is equal to the applied voltage subtracted by the built-in voltage. Using the equilibrium concentrations of $n(x), p(x)$ and the equilibrium internal field strength $E(x)$ as the initial conditions, the continuity equations and the Poisson's equation are evolved together under the constant illumination condition to reach the steady state solutions, from which the $J-V$ curves are plotted. The simulation parameters are presented in Table. \ref{para} except being noted otherwise.

\begin{table}
 \caption{The parameters used in the device model simulation\label{para}}
 \begin{ruledtabular}
 \begin{tabular}{ccc}
 Parameter &  Symbol & Value \\
  \hline
  Donor (Acceptor) band gap & $E_g$ & 1.8\,eV   \\
  Injection barriers & $\phi_n, \phi_p$ & 0.2\,eV \\
  Relative permitivity & $\varepsilon$ & 3.5  \\
  Active layer thickness & $L$   & 200\,nm \\
  Effective density of states & $N_C, N_V$  & $10^{21} \mbox{cm}^{-3}$ \\
  Charge carrier Mobilities & $\mu_n, \mu_p$ & $0.1\,\mbox{cm}^2/\mbox{Vs}$ \\
  CT generation rate & $G$  & $3\times 10^{21} \mbox{cm}^{-3}\mbox{s}^{-1}$ \\
  CT state lifetime & $\tau$ & 100\,ns \\
  CT state radius &  $a$  & 2.25\,nm \\
  Coherent CT rate & $k_0$ & 0.1\,$\mbox{ns}^{-1}$ \\
  Bimolecular recombination reduction factor &  $\zeta$  & 0.1 \\
  Reorganization energy & $\lambda$ & 0.5\,eV \\
 \end{tabular}
 \end{ruledtabular}
\end{table}

\section{Results and discussion}\label{results}
\subsection{Charge separation through the incoherent Marcus mechanism}
\begin{figure}
  \includegraphics[width=8cm]{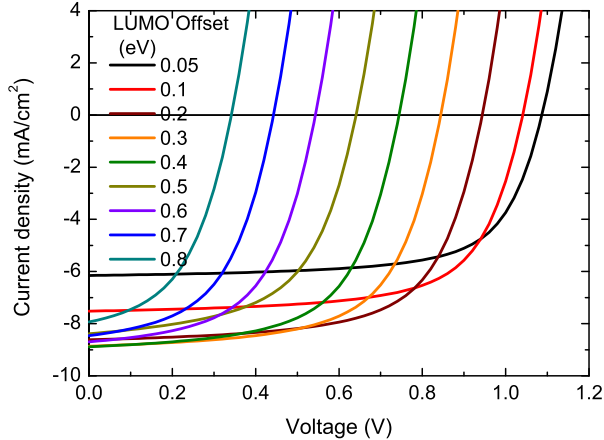}
  \caption{The calculated $J-V$ curves for different interfacial LUMO level offsets (the driving forces) under the Marcus charge separation mechanisms. The reorganization energy $\lambda$ is set to 0.5\,eV. In the calculation the Onsager-Braun theory for CT state dissociation is taken into account. }\label{jv1}
\end{figure}

Based on the assumption that the major role of the driving force is to predominantly determine the charge transfer rate as described by the Marcus theory, we calculated the $J-V$ curves under a set of LUMO level offsets $\Delta E_\mathrm{L}$'s to examine their effects on the device performance, which are shown in Fig.~\ref{jv1}. It is observed that the calculated curves reflect well some features of the experimentally measured $J-V$ curves for the polymer/fullerene bulk heterojunction solar cells with a fixed donor material and varied acceptor materials, that is if a $J-V$ curve shows a high $V_\mathrm{oc}$ the corresponding $J_\mathrm{sc}$ is relatively small, and vice versa\cite{dnuzzo,hoke}. Therefore, it is demonstrated that the excess free energy required for achieving a sufficiently high nonadiabatic charge transfer rate $k_\mathrm{PET}$ could be the probable origin of the observed tradeoff between the $J_\mathrm{sc}$ and $V_\mathrm{oc}$ in these devices. Generally, as the driving force increases by 0.1\,eV each time, the $V_\mathrm{oc}$ decreases exactly by 0.1\,V, which is simply due to the consequent decreasing of the effective band gap. In the following we will mainly focus on the behavior of $J_\mathrm{sc}$ with the varying driving forces. It can be observed that the $J_\mathrm{sc}$ increases significantly with the increasing $\Delta E_\mathrm{L}$ when $\Delta E_\mathrm{L}$ is as small as 0.1 or 0.2\,eV. In this case the enhanced electron transfer rate $k_\mathrm{PET}$ produces high concentration of CT states, whose subsequent dissociation leads to the increased photocurrent. As the $\Delta E_\mathrm{L}$ approaches 0.5\,eV, the $J_\mathrm{sc}$ reaches its maximum and then decreases.

\begin{figure}
  \includegraphics[width=16cm]{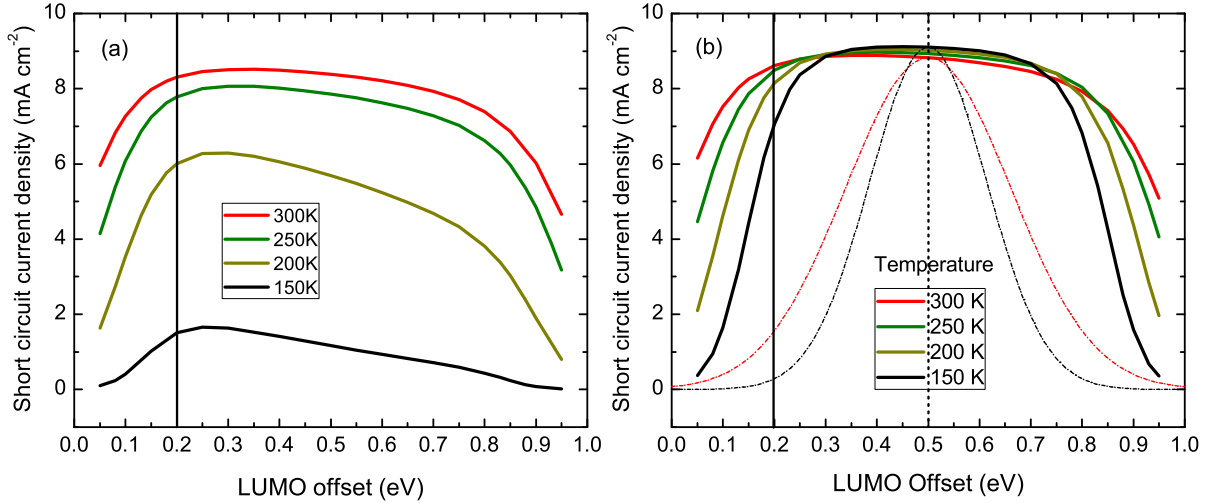}
  \caption{The calculated short circuit current density versus the interfacial LUMO offset ($J_\mathrm{sc}-\Delta E_\mathrm{L}$) under the Marcus charge transfer mechanism at different temperatures, with the field and temperature dependent Onsager-Braun CT state dissociation probability $P(E)$ being taken into account (a) or neglected (b) in the calculation. For comparison, the corresponding $k_\mathrm{PET}-\Delta E_\mathrm{L}$ curves with their maximum being scaled to the value of $J_\mathrm{sc}$ at $\Delta E_\mathrm{L}=\lambda=0.5$\,eV are also plotted in (b). }\label{jsc}
\end{figure}

In order to reveal quantitatively the relationship between the steady state photocurrent and the driving force, we calculated more $J_\mathrm{sc}$'s under different $\Delta E_\mathrm{L}$'s and plotted them in Fig.~\ref{jsc}(a), where the effect of temperature is also examined considering the strong temperature-dependence of $k_\mathrm{PET}$ and $P(E)$. At the room temperature (RT) of 300\,K, as the $\Delta E_\mathrm{L}$ increases the $J_\mathrm{sc}$ quickly rises to over $8\,\mbox{mA}/\mbox{cm}^2$, and keeps this high and approximately constant value in the wide range from 0.2 and 0.7\,eV, beyond which the Marcus inverted region emerges and the $J_\mathrm{sc}$ becomes smaller. With the decreasing temperature, the $J_\mathrm{sc}$ reduces greatly due to the reduction of the dissociation proportion $P(E)$ of the CT states. Moreover, for the curves of low temperature, the high-and-flat region shown in the RT curve disappears, and the $J_\mathrm{sc}$ begins to decrease slowly as soon as it reaches its maximum at $\Delta E_\mathrm{L}=0.3$\,eV. This is because with the increasing $\Delta E_\mathrm{L}$, the built-in field is greatly weakened so that the $P(E)$ decreases as the result of the reduced internal field, leading to inefficient charge extraction and smaller $J_\mathrm{sc}$'s. Therefore, at low temperatures the free charge generation is strongly restricted by the small field dependent CT state dissociation rate.

\begin{figure}
  \includegraphics[width=8cm]{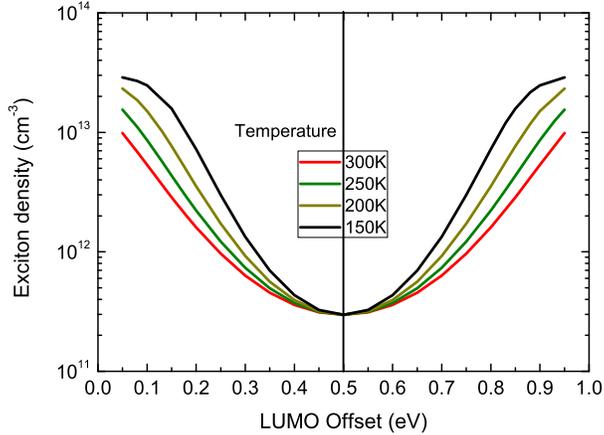}
  \caption{The calculated exciton density versus the interfacial LUMO offset under the Marcus charge transfer mechanism at different temperatures. Each point for density corresponds to the average value over the whole active layer in the device. The reorganization energy $\lambda$ is set to 0.5\,eV}\label{exdis}
\end{figure}

The sole effect of the driving force can be observed in Fig.~\ref{jsc}(b), where we plotted the $J_\mathrm{sc}-\Delta E_\mathrm{L}$ curves calculated by setting $P(E)=1$. In this case all of the $J_\mathrm{sc}-\Delta E_\mathrm{L}$ curves have the high-and-flat region, even though the region gradually shrinks with the decreasing temperatures. In addition, the curves are basically symmetric with respect to the vertical line of $\Delta E_\mathrm{L}=0.5$\,eV, which is the feature of the Marcus charge transfer rate. Nevertheless, they display large discrepancy with the corresponding $k_\mathrm{PET}-\Delta E_\mathrm{L}$ curves, for the latter have prominent peaks when the $\Delta E_\mathrm{L}=\lambda$ and reduces much more rapidly when the $\Delta E_\mathrm{L}$ deviates from the $\lambda$. This result is in contradictory with Coffey et al's finding that the photo-carrier relative yield data measured with the time-resolved microwave photoconductivity (TRMC) method for blends of fixed acceptor and different donors can be well fitted by the $k_\mathrm{PET}-\Delta E_\mathrm{L}$ curve\cite{coffey}. To find the underlying reason of the discrepancy, we calculated the steady state exciton concentration $X$ under the short circuit condition for different $\Delta E_\mathrm{L}$'s and temperatures, which are averaged over the whole active layer thickness, as shown in Fig.~\ref{exdis}. It is seen that the variation of the exciton concentration $X$ with respect to $\Delta E_\mathrm{L}$ is just opposite to that of  $k_\mathrm{PET}$, i.e. the curve has a deep valley precisely at the point of the reorganization energy $\lambda$ of 0.5\,eV, and this feature does not change at different temperatures. Therefore, the product of $k_\mathrm{PET}X$ which is the CT states generation rate is approximately a constant for a wide range of $\Delta E_L$, which is the origin of the high-and-flat region for the $J_\mathrm{sc}-\Delta E_\mathrm{L}$ curves. Especially for small $\Delta E_\mathrm{L}$'s, under the constant illumination condition in steady state the induced small $k_\mathrm{PET}$ may give rise to high concentration of unquenched excitons, and many of which can dissociate within their lifetime to contribute to the photocurrent. On the contrary, under the pulsed illumination condition in the TRMC experiments, since the supply of excitons is limited by the light-pulse duration for each type of polymer:fullerene blend, only the $k_\mathrm{PET}$ basically governs the variational tendency of the photo-carrier yield with respect to the $\Delta E_\mathrm{L}$, such that a bell-like curve for the photo-carrier yield emerges\cite{coffey}.

The occurrence of high $J_\mathrm{sc}$ under small $\Delta E_\mathrm{L}$ in steady state suggests that it is unnecessary to employ pairs of donor and acceptor materials with their $\Delta E_\mathrm{L}$ approaching the reorganization energy $\lambda$ to achieve the maximum photocurrent. According to Fig.~2, at the RT with $\lambda=0.5$\,eV, a moderate $\Delta E_\mathrm{L}$ of 0.2-0.3\,eV can provide sufficient driving force for charge separation at the D/A interface. Thus in principle given a donor material, much $V_\mathrm{oc}$ loss due to the interfacial energy level offset could be saved by employing acceptors with higher LUMO levels to form blend with the donor. However, in some polymer/fullerene blended systems the required driving force for achieving sizable photocurrent is still as high as 0.5\,eV\cite{wright,hendriks}. This may be due to the fact that both of the electron transfer and the hole transfer processes contribute to the photocurrent, and the driving force for the latter is experimentally revealed to be 0.3\,eV higher than the former\cite{hendriks}. Consequently, if a significant proportion of the excitons are dissociated through transferring their holes from the acceptor to the donor, a relatively large HOMO offset $\Delta E_\mathrm{H}$ is essential for achieving a sufficiently high hole transfer rate $k_\mathrm{HT}$ and thus the high $J_\mathrm{sc}$. The situation may occur in devices made of non-fullerene acceptors, in which the photo-absorption of acceptors contribute greatly to the exciton formation\cite{stoltzfus}. Moreover, it is reported that even for fullerene-based acceptors such as ICBA, a large proportion of excitons generated in the polymers can diffuse into them through the F$\ddot{\mbox{o}}$rster resonant energy transfer process, and these excitons can only dissociate through the hole transfer pathway, and the inefficient hole transfer process in polymer-ICBA devices should be responsible for their low $J_\mathrm{sc}$\cite{dnuzzo,hoke}.

\begin{figure}
  \includegraphics[width=8cm]{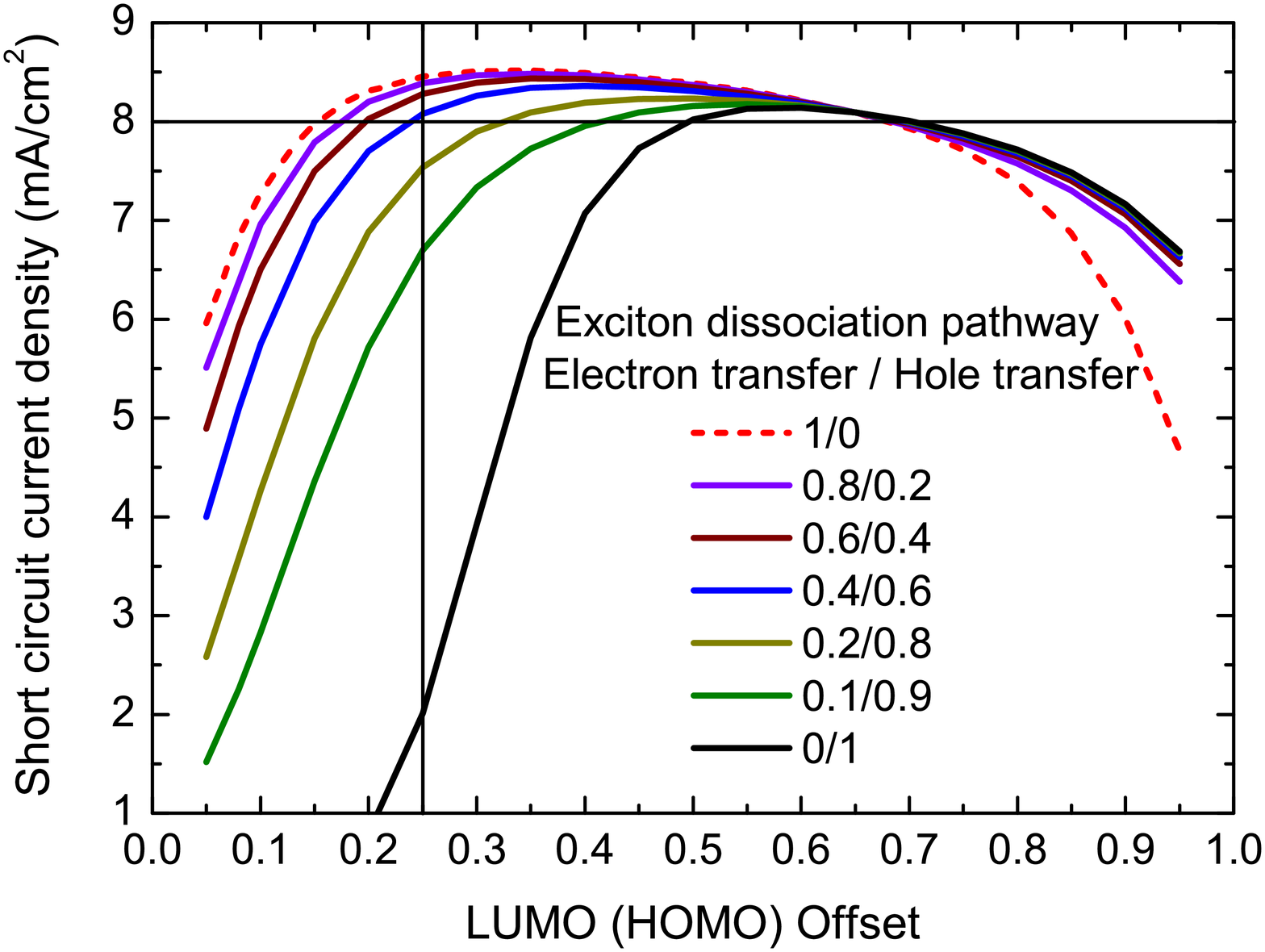}
  \caption{The calculated $J_\mathrm{sc}-\Delta E_\mathrm{L}$ curves under the Marcus charge transfer mechanism for a set of different proportions $P_e/(1-P_e)$ of the electron/hole transfer pathways. The reorganization energy $\lambda_\mathrm{L}, \lambda_\mathrm{H}$ for the electron and hole transfer pathways are set to 0.5 and 0.8\,eV, respectively. }\label{twopathway}
\end{figure}

Here we denoted the respective proportions of excitons dissociating through the two pathways as $P_e$ and $P_h=1-P_e$, and calculated the $J_\mathrm{sc}-\Delta E$ curves with the varied $P_e$ to evaluate the combined roles of the electron and hole transfer pathways on the $J_\mathrm{sc}$. Since we assumed the same band gaps for the donor and acceptor materials, the HOMO offset $\Delta E_\mathrm{H}$ is equal to the LUMO offset $\Delta E_\mathrm{L}$; and the reorganization energy for hole transfer is 0.3\,eV higher than that of the electron transfer. The calculated results are shown in Fig.~\ref{twopathway}. It is observed that as more excitons are generated in or transferred into the acceptor, the high-and-flat region for $J_\mathrm{sc}$ is maintained, but its onset is shifted to the higher $\Delta E$, which suggests that the driving force required for achieving sizable photocurrent becomes larger if the hole transfer reaction, being of a higher reorganization energy, plays a significant role on exciton dissociation. In addition, the slow decreasing of $J_\mathrm{sc}$'s in the high $\Delta E$ regime is mainly caused by the reduced built-in field rather than the Marcus inverted effect, which is different from the sole electron transfer case (the red dashed line).

\begin{figure}
  \includegraphics[width=8cm]{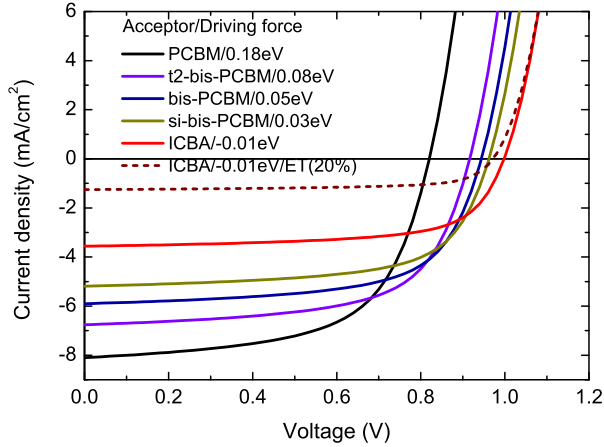}
  \caption{The calculated $J-V$ curves for devices with the donor of PF10TBT blended with different acceptors. The effective energy gap is set to 1.66\,eV; the reorganization energy is set to 0.5\,eV; and the electron transfer driving forces $\Delta E_\mathrm{L}$ are set to 0.18, 0.08, 0.05, 0.03 and 0.01\,eV, corresponding to the acceptors of PCBM, $\mbox{t}_2$-bis-PCBM, bis-PCBM, si-bis-PCBM and ICBA, respectively. The dashed line is calculated for the ICBA acceptor by taking into account the hole transfer pathway with $P_e=0.2$ and $\lambda_\mathrm{H}=0.8$\,eV, which can better fit the experimentally measured $J-V$ curve for the ICBA. }\label{jvicba}
\end{figure}

Based on the above understanding, we calculated the RT $J-V$ curves with the $\Delta E$'s derived from real materials, which are some types of fullerene-based acceptors blended with the donor of PF10TBT, as shown in Fig.~\ref{jvicba}. The effective energy gap is set to 1.66\,eV, and the electron transfer driving forces $\Delta E_\mathrm{L}$ are set to 0.18, 0.08, 0.05, 0.03 and 0.01\,eV, corresponding to the acceptors of PCBM, $\mbox{t}_2$-bis-PCBM, bis-PCBM, si-bis-PCBM and ICBA, respectively. Without taking into account the hole transfer pathway, the high $J_\mathrm{sc}$ exhibited by the curve of PCBM further suggest that only a small driving force of about 0.2\,eV is required for obtaining sufficiently high photocurrent. Compared with the experimentally measured curves in Ref.\cite{dnuzzo}, the curve of ICBA exhibits a much larger $J_\mathrm{sc}$ of 3.56\,$\mbox{mA/cm}^2$. So we speculate that when examining the performance of the ICBA-acceptor devices, it is important to incorporate the effects of the high proportion of excitons diffusing into the ICBA and the inefficient hole transfer pathway. With the $P_e=0.2$ and the hole transfer reorganization energy $\lambda_\mathrm{H}=0.8$\,eV, the recalculated $J-V$ curve (the dashed line) of ICBA gives rise to a relatively realistic $J_\mathrm{sc}$ of 1.25\,$\mbox{mA/cm}^2$. It is noticed that in this case the $V_\mathrm{oc}$ also becomes smaller because of the reduced photo-carrier density under the open circuit condition.

\subsection{Charge separation through the coherent/ballistic dynamics}
Next we examine the role of the driving force in coherent mechanism for charge separation. In contrast to the Onsager-Braun theory, for the coherent mechanism the charge transfer process is ballistic and band-like rather than diffusive, and results in complete charge separation. The relevant exciton dissociation rate can be assumed to be of a constant rate of $k_0$ which is independent of the ambient temperature and the electric field. But as have been demonstrated by lots of experiments like the PPP, only the proportion of excitons that are energetically resonant with the delocalized CT states (thus they can be deemed as the same entity) in the CT states manifold can participate the ballistic transfer process\cite{bakulin1,bakulin2}. To make the picture of the involved energy levels simple, it is assumed that the delocalized CT states accessible for charge separation are those states which are energetically higher than the LUMO level of the acceptor, and the CT states below the acceptor LUMO level are relaxed and localized so that they cannot dissociate successfully but decay to the ground state through geminate recombination. Since the coherent charge transfer process is ultrafast, which occurs within 100\,fs upon photo-excitation and is prior to the relaxation of the hot CT states\cite{grancini,jailaubekov,whaley,yao}, initially the population in each level of the CT states manifolds does not obey the equilibrium distribution and is considered to be evenly distributed. In particular, the population of the in-gap CT states arises from the low energy such as the near infrared photo-absorption\cite{troisi}, as has been verified by some external quantum efficiency (EQE) measurements\cite{vandewal}. According to the above considerations, if the total number of states in the intermediate CT states manifold is fixed to be $N$, the corresponding density of states (DOS) $g(E)$ solely determines the proportion of the delocalized CT states $P_\mathrm{band}$ in the whole manifold, thus we have
\begin{equation}\label{pband}
    P_\mathrm{band}(\Delta)=\frac{1}{N}\int_{E^\prime-\Delta}^{E^\prime} g(E) dE,
\end{equation}
where $E^\prime$ is the upper limit of the CT states manifold and $\Delta$ is the width of the energy window of the delocalized electronic states, i.e. the interfacial LUMO level offset. To investigate the dynamics of free charge carrier, only the delocalized CT states should be taken into account. Whereas the low-lying ones do not contribute to the photocurrent and are just wasted. Thus the continuity equations of excitons should be modified into the form of
\begin{equation}\label{exciton2}
    \frac{\partial X}{\partial t}=D_X\frac{\partial^2 X}{\partial x^2}-\frac{X}{\tau}-k_0 X  + P_\mathrm{band}G.
\end{equation}
Solving Eq.~(\ref{exciton2}) together with other device model equations, the $J-V$ curves for the coherent charge separation mechanism can be obtained.

To obtain the actual $P_\mathrm{band}$, the $g(E)$ must be explicitly given. However, for real materials there are many complicated effects impacting the CT DOS, such as the energetic disorder\cite{mcmahon}, the entropic effect due to the different dimensionality of the donor and acceptor molecules\cite{bregg}, the aggregation effects of the fullerene-based acceptors and the image charge effect at the donor/acceptor interface\cite{savoire,xyz}. Currently the $g(E)$ can only be calculated using first principle or molecular dynamics methods for specific donor-acceptor material systems\cite{tamura,chner,nan,savoire,belj}, and there lacks an analytical expression for it. For the convenience of the present phenomenological investigation, we assumed two simplified analytical expressions of $g(E)$ which may approximate the real DOS of the CT states manifold to some extent.

\subsubsection{Hydrogen-atom-like density of states}
\begin{figure}
  \includegraphics[width=16cm]{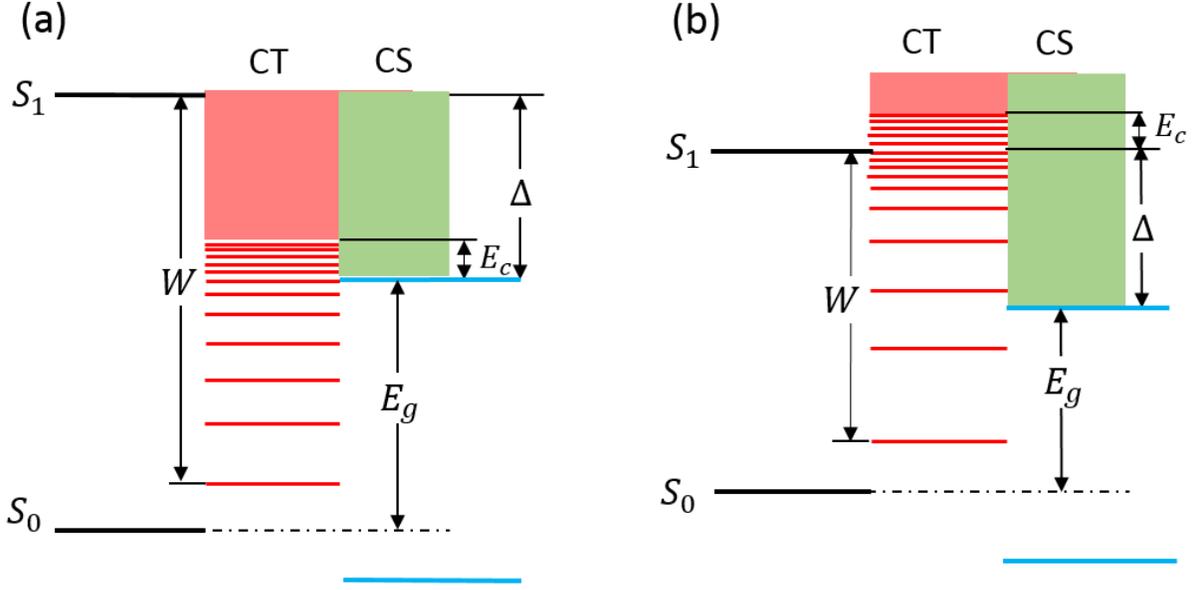}
  \caption{The schematic illustration of the donor/acceptor interfacial energetics which contains the delocalized CT states and may facilitate the coherent (ballistic) charge separation. The energy levels of donor are on the left side and those of acceptor are on the right side. The Charge transfer (CT) states manifold (marked in red) is of the width $W$, in which the levels above the acceptor LUMO level are delocalized and resonant with the charge separated (CS) states (marked in green), forming an energy window of $\Delta$ in which the ballistic charge transfer can take place. The energy parameter $E_c$ represents a critical energy level on which the continuous and the discrete spectra in the CT states manifold meet. The $E_c$ level may be in the energy window (a), or above it (b). The case of (b) may occur in the hydrogen-atom-like DOS of the CT states manifold. The specific forms of the CT states DOS are described in the text. }\label{schematic}
\end{figure}

Firstly, considering the CT excitons as bounded polaron pairs, their energy levels resemble those of a hydrogen atom so that the $g(E)$ has the hydrogen-atom DOS like expression\cite{xyz}. In Fig.~\ref{schematic}(a) we schematically depicted the interfacial energetics, where the zero point of the CT state energy is set to be the acceptor LUMO level, i.e. $E_\mathrm{ct}$ is transformed to $E_\mathrm{ct}-E_g$. A cutoff energy level $E_c$ slightly above the acceptor LUMO level separates the high-lying continuous energy spectrum from the low-lying discrete one. Then for $E\geq E_c$, $g(E)$ equals to a constant of $\alpha (E_a-E_c)^{-3/2}$; while for $E<E_c$, $g(E)=\alpha (E_a-E)^{-3/2}$, in which the parameter $E_a$ is a small positive energy used to avoid the singularity in $g(E)$, and the prefactor $\alpha$ is determined by the normalization condition of
\begin{equation}\label{normalization}
    N=\int_{E^\prime-W}^{E^\prime} g(E) dE
\end{equation}
with $W$ the width of the CT state manifold. Based on the Eqs.~(\ref{pband},\ref{normalization}), the proportion $P_\mathrm{band}$ is deduced to be of the form of
\begin{equation}\label{pbandhy}
   P_\mathrm{band}(\Delta)=\frac{(\Delta-E_c)/(E_a-E_c)+2\left[1-\sqrt{(E_a-E_c)/E_c}\right]}{(\Delta-E_c)/(E_a-E_c)+2\left[1-\sqrt{(E_a-E_c)/(E_c+W-\Delta)}\right]}.
\end{equation}

\begin{figure}
  \includegraphics[width=8cm]{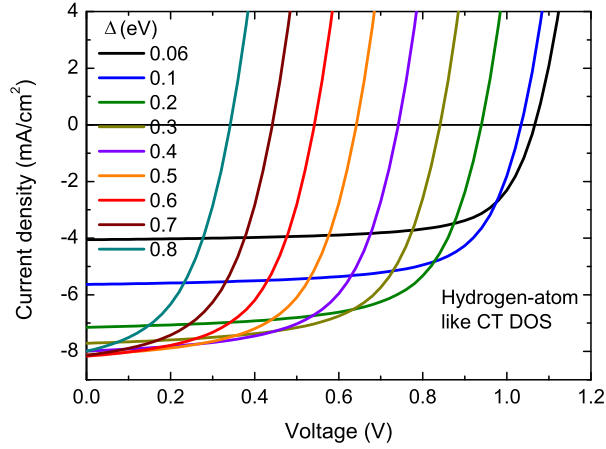}
  \caption{The calculated $J-V$ curves for different driving force $\Delta$'s under the coherent charge separation mechanism. The CT states manifold is of a hydrogen-atom-like DOS, with the energy parameter $E_c=0.05$\,eV, $E_a=0.1$\,eV and $W=1$\,eV. }\label{jvpband}
\end{figure}

Substituting the $P_\mathrm{band}$ into Eq.~(\ref{exciton2}) with $E_c=0.05$\,eV, $E_a=0.1$\,eV and $W=1$\,eV, we calculated the $J-V$ curves with a set of different driving force $\Delta$'s, which are plotted in Fig.~\ref{jvpband}. It is observed that the curves exhibit the exactly same behavior as those calculated for the Marcus incoherent charge transfer mechanism (Fig.~\ref{jv1}). That is with the increased $\Delta$, the $J_\mathrm{sc}$ increases whereas the $V_\mathrm{oc}$ decreases evenly. Thus it is not plausible to gain some clues concerning which charge separation mechanism is the dominant one just from the variation of $J-V$ curves with respect to the driving force. For $\Delta=0.2$\,eV, the optimum device performance is obtained with $J_\mathrm{sc}=7.14\,\mbox{mA/cm}^2$ and $V_\mathrm{oc}=0.94$\,V, such that a driving force of 0.2\,eV is sufficient for achieving balanced $J_\mathrm{sc}$ and $V_\mathrm{oc}$ in devices where the coherent exciton dissociation mechanism plays the major role on charge generation.

The quantitative relationship between the $J_\mathrm{sc}$ and $\Delta$ are calculated and presented with the varying cutoff energy $E_c$ and the fixed $E_a=0.1$\,eV, as shown in Fig.~8(a). For all the curves, the $J_\mathrm{sc}$ increases very rapidly with $\Delta$ when the latter is smaller than 0.2\,eV, because in the energy window of 0-0.2\,eV the CT states form a quasi-continuum band with high DOS $g(E)$, and a small increase of $\Delta$ can induce high extra population of the CT states to participate the coherent (ballistic) charge transfer and separation. For $E_c=0.08$\,eV, the $J_\mathrm{sc}$ reaches an approximately constant high value beyond $\Delta=0.2$\,eV, being similar to the high-and-flat region appearing in the $J_\mathrm{sc}-\Delta E_L$ curve under the incoherent charge transfer mechanism. On the other hand, with the reduced $E_c$ the $J_\mathrm{sc}$ becomes smaller and remain increases slowly in the relatively high $\Delta$ regime, which is due to the fact that the in-gap CT states close to the acceptor LUMO level are of a high DOS and thus need to be harvested by enhancing the $\Delta$ in order to reach a sufficiently high $J_\mathrm{sc}$. The decreasing of $J_\mathrm{sc}$ around $\Delta=0.8$\,eV is only caused by the reduction of the built-in electric field and the lowered charge extraction efficiency, rather than the Marcus inverted region in Fig.~2. We also included a curve with the varied $E_a$ of 0.2\,eV and $E_c=0.08$\,eV(the dashed line), where compared to the corresponding curve with $E_a=0.1$\,eV (the red line), the $J_\mathrm{sc}$ reduces significantly as a result of the reduced DOS for the high-lying levels in CT state manifold.

The CT states lying in the energy window $\Delta$ can also possibly consist of purely discrete spectrum, namely the $E_c$ level is equal to or above the donor LUMO level, as shown in Fig.~\ref{schematic}(b). In this case we assumed the $E_c$ and $E_a$ to be the same. Then the delocalized CT states proportion $P_{band}$ is modified to
\begin{equation}\label{pband2}
   P^{\prime}_\mathrm{band}(\Delta)=\frac{1-(1+\Delta/E_c)^{-1/2}}{1-(1+W/E_c)^{-1/2}}.
\end{equation}
Inserting the $P'_\mathrm{band}$ into the device model, the $J_\mathrm{sc}-\Delta$ curves are calculated and presented in Fig.~8(b). The shape of the curves does not exhibit observable change as compared to that in Fig.~8(a), but it becomes more difficult to achieve a sizable $J_\mathrm{sc}$ for the small $\Delta$. Generally the $J_\mathrm{sc}$ increases with the decreasing $E_c$. Therefore the $E_c$ level should be tuned as close to the donor LUMO level as possible. Combined with the results in Fig.~8(a), we conclude that the criterion for good interfacial energetics being able to facilitate significant coherent exciton dissociation is that, the CT states manifold is low and the energy window $\Delta$ is resonant with at least part of the continuous spectrum and the high-lying discrete spectrum, so that plenty of hot excitons can be harvested.

\begin{figure}
  \includegraphics[width=16cm]{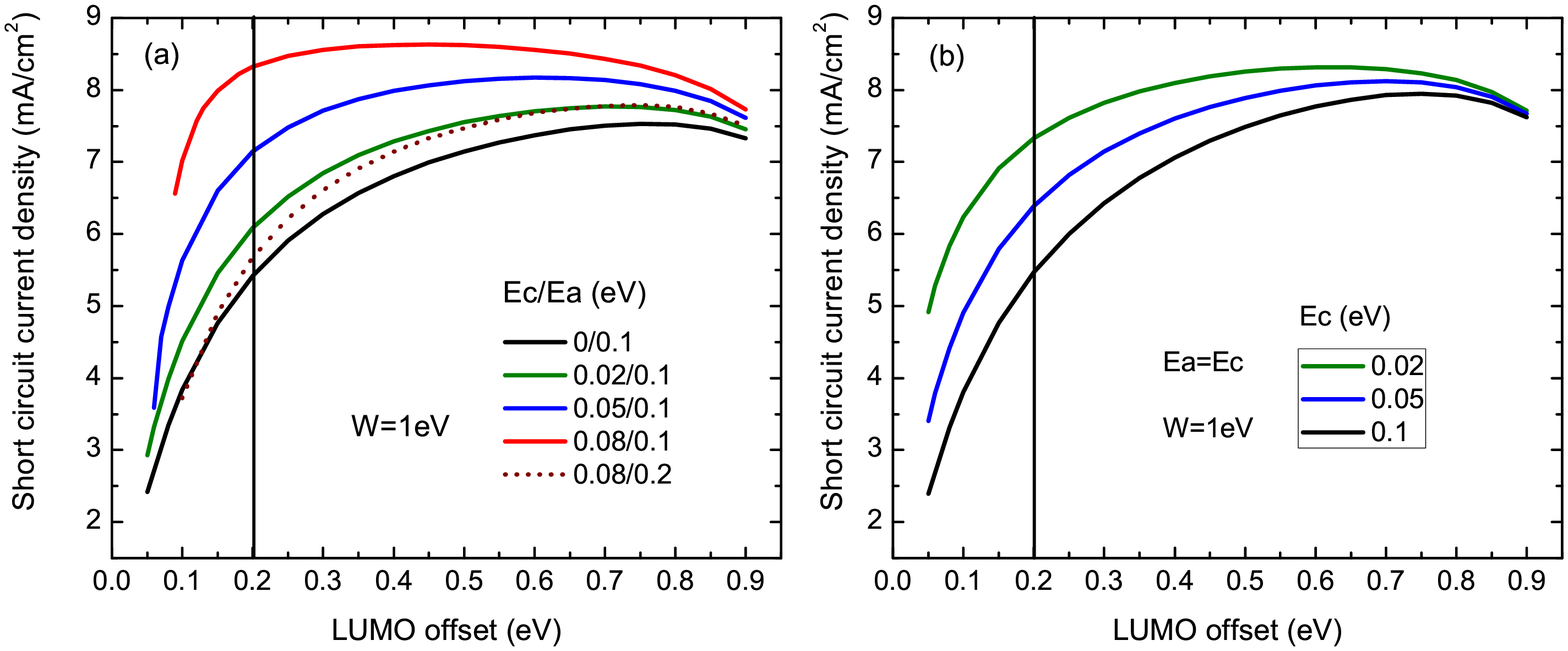}
  \caption{The calculated short circuit current density versus the driving force ($J_\mathrm{sc}-\Delta$) curves for different energy parameters of $E_c,E_a$ of the hydrogen-atom-like CT states DOS. (a) The CT states manifold may consist of the continuous and discrete spectra, corresponding to the DOS schematically illustrated in Fig.~\ref{schematic}(a); (b) or the manifold is of the purely discrete energy levels below the LUMO level of the donor, corresponding to the DOS schematically illustrated in Fig.~\ref{schematic}(b). }\label{jscecea}
\end{figure}

\subsubsection{Exponential density of states}

Secondly, motivated by the mobility edge model in organic semiconductors\cite{neher}, we considered CT states manifold to be of the exponential type of DOS. The interfacial energetic levels can be still basically schematically illustrated by the Fig.~\ref{schematic}(a), with $E_c$ representing the cutoff energy level separating the continuous band from the discrete levels. For $E\geq E_c$, the DOS $g(E)=\alpha/E_a$; while for $E<E_c$, $g(E)=\alpha/E_a \exp[(E-E_c)]/E_a$, where $E_a$ is a parameter characterizing the width of the exponential states, and the prefactor $\alpha$ is also determined by the normalization condition Eq.~(\ref{normalization}). Now it can be deduced that the proportion of the CT states lying in the energy window of the driving force $\Delta$ is
\begin{equation}\label{pbandexp}
     P^{\prime\prime}_\mathrm{band}(\Delta)=\frac{(\Delta-E_c)+E_a[1-\exp(-E_c/E_a)]}{(\Delta-E_c)+E_a\left[1-\exp\left(\frac{\Delta-W-E_c}{E_a}\right)\right]}.
\end{equation}

\begin{figure}
  \includegraphics[width=16cm]{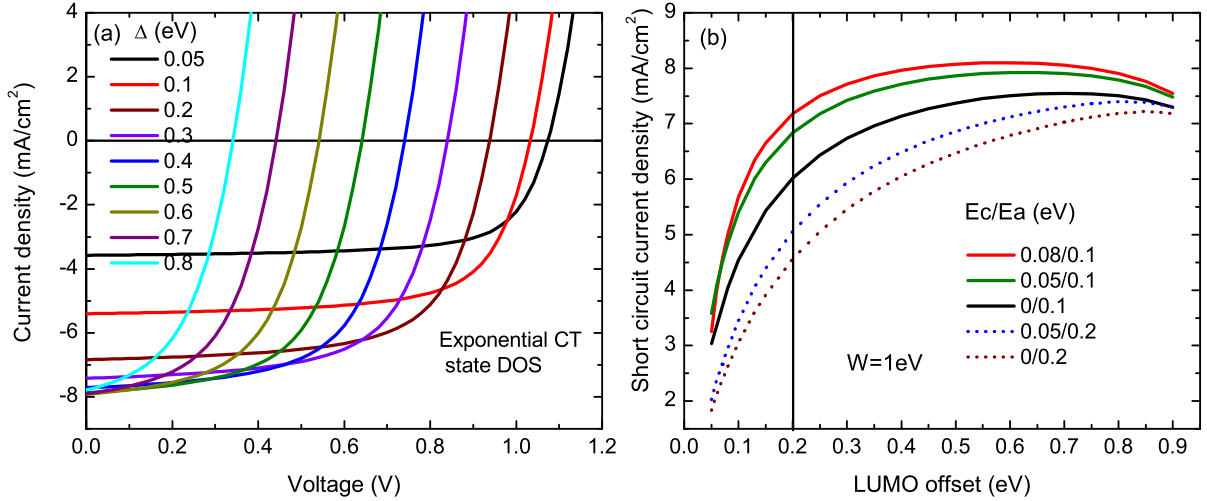}
  \caption{(a) The calculated calculated $J-V$ curves for different driving force $\Delta$'s under the coherent charge separation mechanism. The CT states manifold is of an exponential DOS, with the energy parameter $E_c=0.05$\,eV, $E_a=0.1$\,eV and $W=1$\,eV. (b) The calculated $J_\mathrm{sc}-\Delta$ curves for different energy parameters of $E_c,E_a$ of the exponential CT states DOS, as schematically illustrated in Fig.~\ref{schematic}(a). }\label{expjv}
\end{figure}

With the energy parameters $E_c, E_a$ and $W$ being set to 0.05, 0.1 and 1\,eV, we substituted the above $P^{\prime\prime}_\mathrm{band}(\Delta)$ into Eq.~(\ref{exciton2}) and calculated the $J-V$ curves for different driving force $\Delta$'s, as shown in Fig.~\ref{expjv}(a). It is obvious that the curves show almost the same features with respect to those for hydrogen-atom like CT state DOS (see Fig.\ref{jvpband}), suggesting that the device performance is not very sensitive to the specific form of the DOS as long as the number of in-gap levels decreases quickly with the decreasing energy. In Fig.~\ref{expjv}(b) we present the calculated $J_\mathrm{sc}-\Delta$ curves for the varying $E_c$ and $E_a$. Similar to the behaviors exhibited by the curves for hydrogen-atom like DOS, the relatively larger $E_c$ more or less give rises to the higher $J_\mathrm{sc}$, because of the inclusion of the dense high-lying discrete levels in the energy window for ballistic charge transfer. On the other hand, the $E_a$ plays a much more important role on determining the photocurrent. With the increasing of $E_a$ from 0.1\,eV to 0.2\,eV, the $J_\mathrm{sc}$ decreases by nearly 2\,$\mbox{mA/cm}^2$ under the $\Delta$ of 0.2\,eV. In order to obtain a sizable $J_\mathrm{sc}$ under a small driving force, the width of the in-gap states in the CT manifold should be restricted to a value at least being smaller than 0.2\,eV. Therefore, the optimization of the DOS for the CT sates manifold is important, which could be realized by changing donor:acceptor ratio of the blend to enhance the fullerene aggregation and crystallization so that more delocalized CT states may be formed.

\section{Conclusion and outlook}\label{conclusion}

In this work, employing the phenomenological device model method we investigated the impacts of the charge separation driving force, which is defined as the donor/acceptor interfacial energy level offsets on the device performance of organic bulk heterojunction solar cells. The driving force $\Delta$ may either provide the free energy required for the incoherent Marcus charge transfer processes to happen or form an energy window where the delocalized CT states reside and facilitate the coherent charge transfer processes. Both of the two kinds of charge separation mechanisms probably play important roles and thus were studied independently by calculating the corresponding $J-V$ and $J_\mathrm{sc}-\Delta$ curves. Generally the $V_\mathrm{oc}$ reduces evenly with the increased $\Delta$, forming a significant $V_\mathrm{oc}$ loss pathway. For the Marcus charge transfer mechanism, with the increasing of $\Delta$ from 0\,eV, the $J_\mathrm{sc}$ initially increases extremely rapidly and begin to saturate under a small delta of 0.2\,eV or so; then the $J_\mathrm{sc}$ maintains a high and nearly constant value until the Marcus inverted effect emerges under too high $\Delta$'s, exhibiting a behavior which is largely different from that of the Marcus charge transfer rate $k_\mathrm{PET}$. The underlying reason is found that the reduced $k_\mathrm{PET}$ under a $\Delta$ deviating from the reorganization energy $\lambda$ is precisely compensated by the enhanced density of the accumulated exciton within their lifetime, such that the overall free charge generation rate changes very slowly. When the hole transfer pathway plays innegligible roles on charge separation, the required $\Delta$ for obtaining a sizable $J_\mathrm{sc}$ may become higher due to the relatively larger reorganization energy on the acceptor side, such as the case for the ICBA acceptor based devices.

For the coherent mechanism, when calculating the $J-V$ and $J_\mathrm{sc}-\Delta$ curves we assumed the hydrogen-atom-like DOS and the exponential DOS for the interfacial CT states manifold, respectively. The results show similar behaviors and suggest that as long as the energy window formed by the interfacial energy offset (or the driving force) contains part of the continuous  spectrum and the dense high-lying discrete levels in the CT state manifold while the low-lying in-gap levels are rare, a great proportion of the CT states can be converted into the fully separated charge carriers and consequently the high $J_\mathrm{sc}$ is obtained under a small $\Delta$ of about 0.2\,eV, which is consistent with the behavior of $J_\mathrm{sc}$ calculated for the incoherent mechanism. Therefore, regardless of the charge separation mechanism, people can obtain the relatively high $J_\mathrm{sc}$ and $V_\mathrm{oc}$ simultaneously without sacrificing one for the other, which may be hopefully realized in the recently popular non-fullerene acceptor solar cells.

In addition, concerning the concrete charge separation mechanism in the actual donor/acceptor blended systems, the coherent and incoherent mechanisms may coexist, which is probably the reason that up to now, in different experiments people have observed that the photocurrent generation follows both the Marcus-type behavior with respect to the driving force and the composition dependence on the donor:acceptor blend ratio. It is demonstrated in our simulation that with a moderate driving force, there is no obvious feature on the $J-V$ curves and the $J_\mathrm{sc}-\Delta$ curves that can identify which one is the dominant mechanism. However, the incoherent mechanism induces strongly temperature-dependent effects for the photocurrent and thus can be singled out through observing the behavior of $J_\mathrm{sc}$ at the lowered ambient temperature. Also, future works on the DOS of CT states may be helpful for acquiring the high $J_\mathrm{sc}$ under the smaller driving forces.

\begin{acknowledgments}
The authors would like to thank Professor R. $\ddot{\mbox{O}}$sterbacka for the fruitful discussion and his insightful comments. This work is supported by the National Natural Science Foundation of China under the Contract No. 11604280 and 51602276.
\end{acknowledgments}

\bibliography{dfpaper}

\end{document}